\documentclass[12pt,notitlepage]{revtex4-1}

\usepackage{graphicx}	% Include figure files
\usepackage{xspace}     % Include xspace
\usepackage{amsmath}
\usepackage{subcaption}

\begin{document}

\title{Cosmic Ray Muon Radiography Applications in Safeguards and Arms Control}

\newcommand{\losalamos}{Los Alamos National Laboratory, Los Alamos, New Mexico 87545, USA}

\affiliation{\losalamos}

\author{J. Matthew Durham} \email[Email address: ]{durham@lanl.gov} \affiliation{\losalamos} 
\date{\today}

\begin{abstract}

Muons are the most penetrating radiographic probe that exists today. These elementary particles possess a unique combination of physical properties that allows them to pass through dense, heavily shielded objects that are opaque to typical photon/neutron probes, and emerge with useful radiographic information on the object's internal substructure. Interactions of cosmic rays in the Earth's upper atmosphere provide a constant, natural source of muons that can be used for passive interrogation, eliminating the need for artificial sources of radiation. These proceedings discuss specific applications of muon radiography in nuclear safeguards and arms control treaty verification.
\end{abstract}

\maketitle

\section{Introduction}

	Highly energetic radiation produced in astrophysical processes is constantly bombarding the Earth.  These naturally occurring cosmic rays provide a window into the dynamics of distant events in our universe, and as such have been the subject of intense scrutiny.  Among recent work, measurements have shown that the incident high-energy cosmic ray flux includes neutrinos \cite{ICECUBE}, gamma rays \cite{FERMILAT,HAWC}, positrons and electrons \cite{AMS}, and heavy nuclei \cite{AUGER,HIRES}.  These particles can have energies much higher than can produced in terrestrial accelerators: iron nuclei with energies above $10^5$ GeV have been observed (see \cite{PDG} for a review).  When these heavy, energetic cosmic ray nuclei collide with nitrogen or oxygen nuclei that are present in the atmosphere, a short-lived phase of quark-gluon plasma is created, which immediately freezes out and produces hundreds of charged pions.  Charged pions have a lifetime of 2.6$\times10^{-8}$ s and have a nearly 100$\%$ branching ratio to decay into a neutrino and a muon.
	
	The cosmic ray muons produced from these decays inherit a significant relativistic boost from their parent particles: the combination of their relatively long lifetime of 2.2 $\mu$s and the effects of time dilation allow enough survive their journey to generate a flux of $\sim$$10^4$ muons per square meter per minute at sea level.  The angular distribution of cosmic ray muons on the surface is roughly proportional to $cos^2\theta_z$, where $\theta_z$ is the angle from the zenith, giving a maximum flux moving vertically and a minimum (though non-zero) flux moving horizontally \cite{Reyna}.  Their energy spectrum is broad, and has has an average of $\sim$4 GeV.  
	
	Muons are charged leptons with a relatively large mass of 105.6 MeV/$c^2$, which gives them a unique set of interactions with matter.  Since they are charged they interact with nuclei via the Coulomb force and undergo multiple scattering while passing through matter \cite{scat1,scat2}.  Negative muons can be captured into atomic orbitals where they may interact with the nucleus and induce subsequent particle emission \cite{MuonCapture}.  Like all leptons, they do not participate in the strong nuclear interaction, and radiative energy loss from $bremsstrahlung$ emission is suppressed by their large mass.  This gives most cosmic ray muons a Bethe-Bloch stopping power near the near the minimum of $-<dE/dx> \approx$ 1 MeV/g/cm$^2$.  These properties allow muons to penetrate large amounts of dense material that absorb photons, electrons, and hadrons of similar energies.

	The uniquely penetrating properties of muons, along with their high energies and ubiquitous availability, make them an attractive probe for radiographic imaging.  The method of muon multiple scattering tomography was invented at Los Alamos National Laboratory in 2003 \cite{Nature, Pried, Schultz}, and is well-suited for targets that are too thick for traditional imaging with x-rays, but are thin relative to targets of muon transmission imaging (mountains, pyramids, etc).  Another method, also under investigation at LANL, uses secondary neutrons produced by muon induced fission to tag and image actinides \cite{MIF}.  In these proceedings we will discuss two specific applications of muon imaging relevant to safeguards and arms control.
	
\section{Fuel Cask Verification}

Plutonium is produced by irradiating uranium fuel in a nuclear reactor.  Spent nuclear reactor fuel spends several years in a cooling pool while short lived fission product decay.  After removal from the pool, and in the absence of reprocessing or permanent storage (or while awaiting transport to such facilities) the spent fuel is typically housed in large dry storage casks \cite{NUREG}.  A typical cask for pressurized-water reactor fuel holds 20 to 30 assemblies in a central basket surrounded by neutron and gamma ray shielding.  This heavy shielding built into the cask, as well as self-shielding between neighboring assemblies, protects facility personnel and the public from the potentially dangerous levels of radiation emitted by the fuel.
	
	Safeguards inspectors maintain continuity-of-knowledge (CoK) of spent fuel assemblies inside casks by applying tamper-indicating seals to cask lids.  However, in cases where the seals fail or are made inaccessible for checks, the only way to reassert CoK is to move the cask back to a cooling pool where it can be safely opened and inspected.  This procedure is expensive and disruptive to operations at the nuclear facility under inspection, and in many cases is not feasible.  Therefore, a standalone \textit{in situ} technique to independently reassert knowledge of the cask's contents is necessary.
	
	Previous work towards achieving this goal has focused on using radiation which escapes the casks to attempt to verify the contents.  These measurements of gamma rays and neutrons can show that a cask contains radioactive material, however, the scattered radiation which passes through the cask's shielding does not possess sufficient information to quantify the number of fuel assemblies inside the cask \cite{Ziock}. This is due to the heavy shielding that is necessary to safely store the spent fuel; the casks work as designed.  Other methods have attempted to measure a cask's radiation signature at one point in time to serve as a ``fingerprint" for comparison at later dates \cite{Ziock2,DSVD}.  However, in the decades that may pass between measurements, there can be significant changes to the cask radiation signature due to decays of fission products in the fuel, which must be corrected for.  Like all methods using emanated radiation, fingerprint comparisons must also correct for backgrounds from neighboring casks at spent fuel storage installations.  These corrections can lead to significant systematic uncertainties on the fingerprint that can limit the discriminatory power of such measurements. 
	
Cask verification is an ideal candidate for application of muon scattering radiography.  Muons are highly penetrating, and can pass through the cask and sample the entire fuel basket volume as they are not significantly attenuated by the cask body or fuel assemblies.  Muons are also an external probe that is not affected by backgrounds from neighboring casks.  The casks are large, static objects, and do not move during the time which cosmic ray muon statistics are acquired.  Since muon scattering is not especially sensitive to fuel burnup, there is little change in the measured muon scattering through a cask as fission products decay, so the previous history of the fuel does not need to be known from operator declarations or require any corrections.  There has been a large amount of work in simulation showing the potential usefulness of the technique for monitoring encapsulated waste \cite{Gustafsson, Jonkmans, Furlan, Liao, Ambrosino, Clarkson2, Frazao, Chatzi2, Boniface, Checchia, Liu}.  A group from Los Alamos and Idaho National Laboratories has recently performed the first definitive measurement of muon scattering in a fuel cask \cite{casks_PRApp}, which proves the technique is sensitive to the fuel content of the cask.

\begin{figure}[h]
\begin{subfigure}{0.4\textwidth}
\includegraphics[scale=0.6]{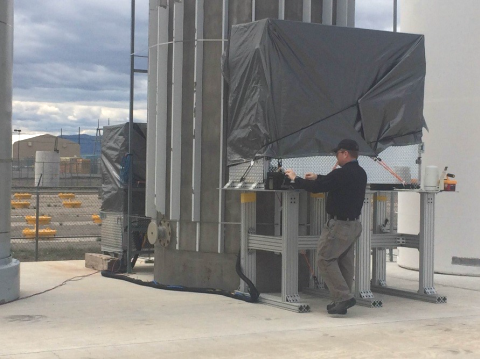}
\label{fig:lightened}
\end{subfigure}
\begin{subfigure}{0.4\textwidth}
\includegraphics[scale=0.25]{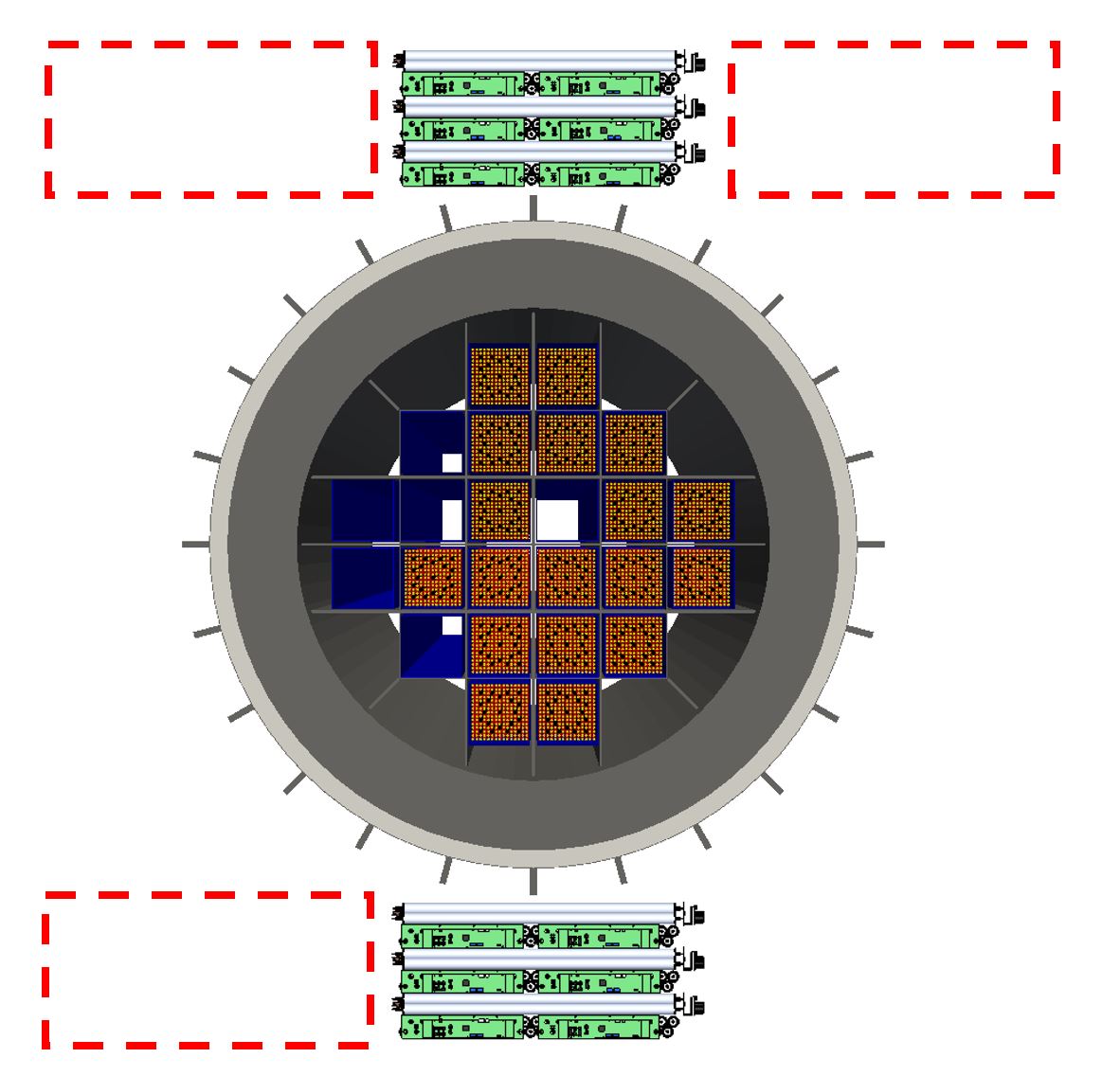}
\label{fig:zebra}
\end{subfigure}
\caption{Left: The two muon trackers around the MC-10 cask at Idaho National Lab.  Right: Top view of the cask loading and measurement positions of the muon tracking detectors. Muons moving between the two detectors pass through columns in the fuel basket containing (from left to right) zero, one, six, five, four, and two fuel assemblies \cite{casks_PRApp}.}
\label{fig:one}
\end{figure}

	The measurement took place at Idaho National Laboratory.  The cask under investigation was a partially loaded metallic-type dry storage cask model MC-10 \cite{MC10} which contained 18 spent PWR fuel assemblies out of a total of 24 slots in the fuel basket.  Identical 1.2$\times$1.2 m$^2$ muon tracking detectors were placed on opposite sides of the cask, and muon scattering data was recorded over a period of $\sim$9 weeks (see Fig. \ref{fig:one}), during which the detectors were moved several times to cover most of the fuel basket.  In each of these positions, between 40 and 90 thousand muon tracks were recorded.  A trigger was implemented in the detector firmware, which requires hits in neighboring tubes within a coincidence window of 600 ns.  This allows muons to be tracked while rejecting backgrounds from signals in single tubes induced by gamma rays or neutrons which escape the cask shielding.

\begin{figure}[h]
\begin{subfigure}{0.4\textwidth}
\includegraphics[scale=0.25]{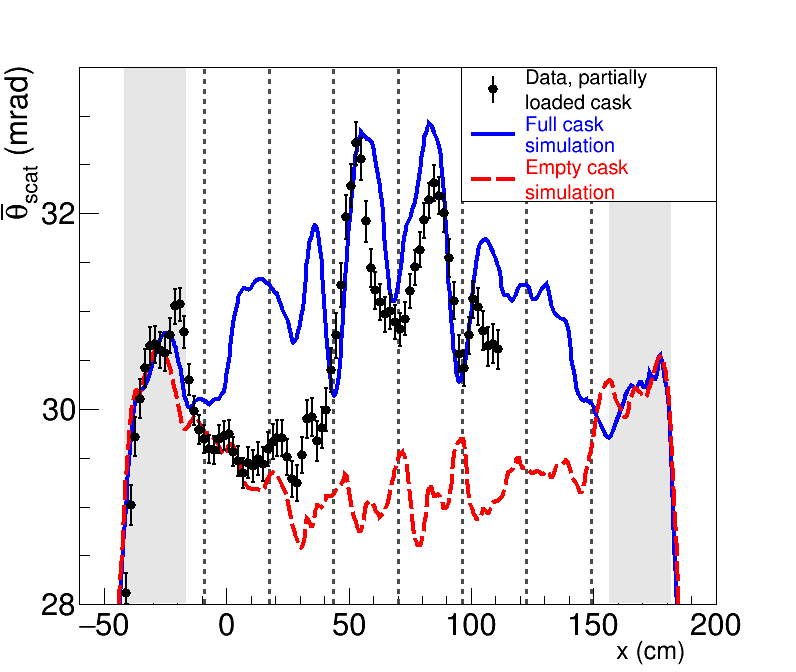}
\label{fig:lightened}
\end{subfigure}
\begin{subfigure}{0.4\textwidth}
\includegraphics[scale=0.25]{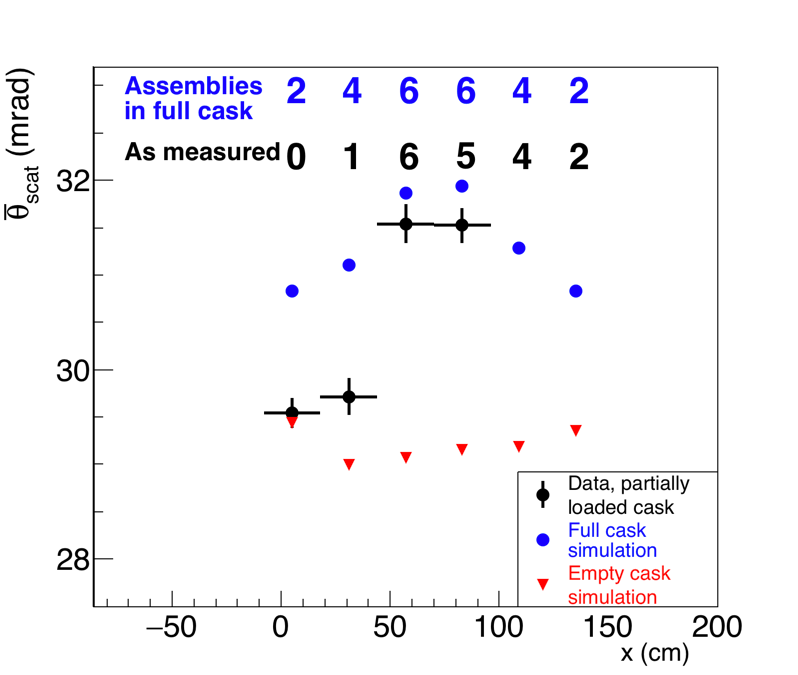}
\label{fig:fig4}
\end{subfigure}
\caption{Left: The average scattering angle across the cask, along with simulations of scattering in a full and empty cask.  Right: The scattering angles integrated over each column in the fuel basket \cite{casks_PRApp}.}
\label{fig:two}
\end{figure}

	  Muon tracks recorded in the upper detector were projected to the center of the cask, where an array of 2 cm wide voxels with corresponding histograms collected the scattering angle of each muon that passed through the voxel.  The average scattering angle for the ensemble of muons passing through each voxel is shown in the left panel of Fig. \ref{fig:two} as a function of horizontal position across the cask.  Also shown is a GEANT4 model \cite{GEANT4} of muon scattering in a full and empty cask, that was constructed with simulated detectors tuned to match the acceptance, efficiency, and angular resolution of the real detectors used in the measurement, and was analyzed with identical techniques.  Weights were applied to the simulated data samples in order to accurately represent the inefficiencies in the detector and different sample sizes in each position.  As can be seen in the simulation of the empty cask, these effects induce non-uniformities in the data that vary across the cask.  The shaded grey areas show the steel shielding in the cask, and the boundaries between columns in the fuel basket are denoted by dashed lines.  The difference between the measurement of the partially loaded cask and the simulation of the full cask due to missing fuel elements is apparent.  Data was also recorded on the right side of the cask, but strong winds at the measurement site shook the detectors during this part of the measurement, which induced artificial angles between incoming and outgoing muon tracks in the data that is therefore not included in this analysis.  This can be corrected in future measurements with more secure detector footing.

For quantitative comparisons between the rows, the muon scattering data is averaged over each column in the fuel basket, from both data and the simulation, and is shown in the right panel of Fig. \ref{fig:two}.  We see that in the two leftmost columns, where there are multiple missing fuel assemblies, there is a difference of more than 5$\sigma$ between the measurement and the expectations of scattering in a fully loaded cask.  This demonstrates that the removal of two or more fuel assemblies can be detected with high confidence.  The full column does show some deviation from expectations at the $<$2$\sigma$ level, which has limited statistical significance, but is not unexpected since muons which pass through this row may also sample partial path lengths through the incompletely filled columns on the left and right.  The column with a single missing fuel assembly shows a deviation from the full cask with a significance of 2.3$\sigma$, which taken alone gives a conclusion at the 98$\%$ confidence level that this row is not fully populated.  As this measurement is statistics limited, a longer measurement time on this column, or the use of larger detectors (which sample a large muon flux) could provide reduced uncertainties on the measured muon scattering and therefore stronger conclusions on the cask contents.  Additional work investigating advanced image reconstruction techniques with computed tomography algorithms \cite{Poulson} and more complicated diversion scenarios is underway.

\section{Treaty Verification}

The procedure of verifying whether a declared treaty regulated object is or is not an active nuclear warhead has an inherent conflict: there must be an unequivocal determination made as to whether or not the presented item is an actual warhead, but the sensitive design details of the object must not be revealed to treaty partners or international inspectors.  Measurements of neutron and gamma rays emitted by an object can provide strong indications of its composition, but may also reveal protected information.  Conventional radiography can provide precise images of the object, which could also raise concerns, and also applies an artificial radiation dose which may be objectionable to treaty partners (see \cite{VerificationReview} for a review of potential techniques).  Muon scattering radiography can provide tomographic images of nuclear warheads which also reveal design details. 

\begin{figure}[h]
	\centering
		 \includegraphics[width=0.9\textwidth]{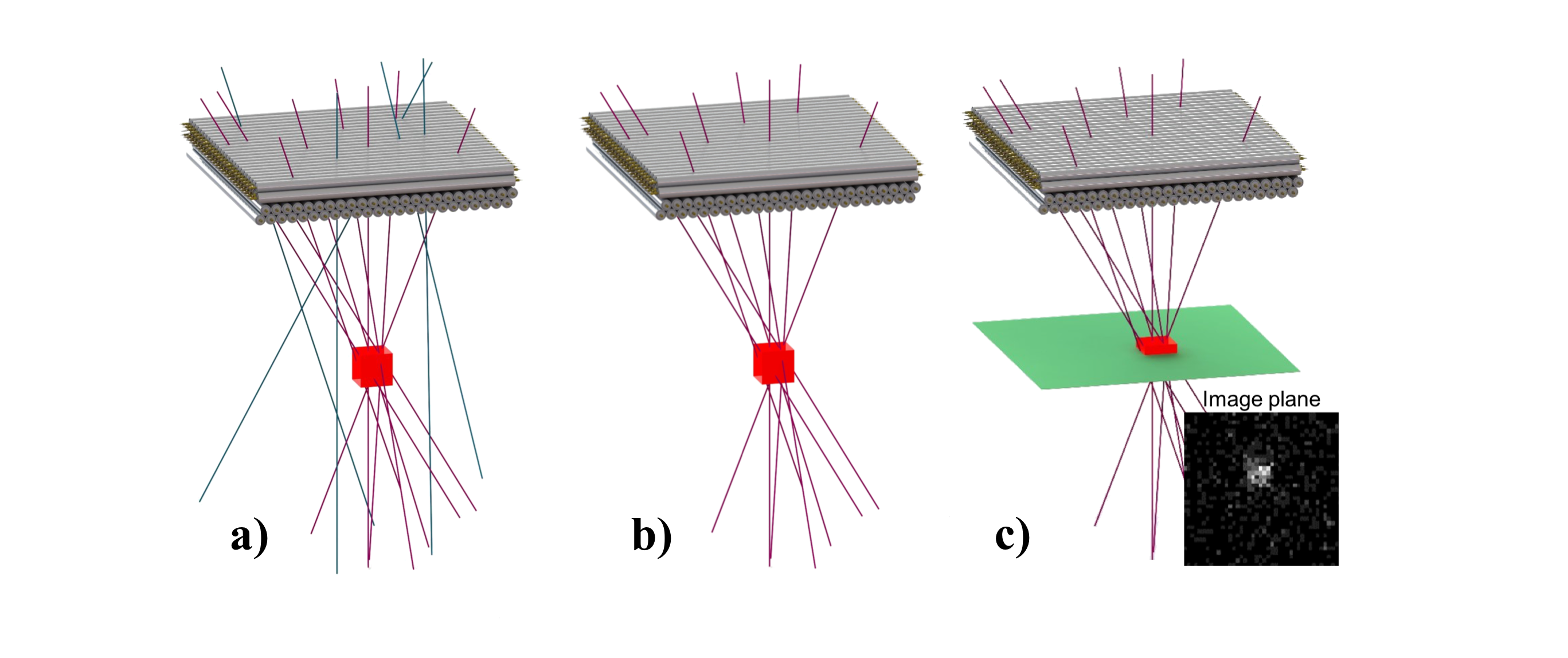}
	\caption{Diagram of imaging with neutron-tagged muons: a) Muons pass through the detector, some of which also pass through the object of interest. b) Muons that induce neutron emission from the object are tagged while others are discarded from the analysis. c) Laminography is performed with the neutron-tagged tracks.  The object comes into focus when the image plane is at the same position as the emission source.} 
	\label{fig:Picture1}
\end{figure}

An alternative method of imaging fissile material uses muons which induce neutron emission from the object under inspection \cite{MIF}.  Negative muons that are stopped in fissile material and captured in atomic orbitals can undergo interactions with the nucleus that lead to subsequent neutron emission \cite{MuonCapture}.  The lifetimes of these muonic atoms are characteristic to each isotope, and have been measured to be 71.6$\pm$0.6 ns for $^{235}$U and 77.2$\pm$0.4 ns for $^{238}$U  \cite{Ulifetime}. With a combination of muon tracking detectors and fast neutron detectors, muons which encounter fissile material can be identified by detecting the emitted neutrons.  An image of the neutron emission source can be made by performing laminography with the neutron-tagged muon tracks.  A diagram of the technique is shown in Fig. \ref{fig:Picture1}.  
 
\begin{figure}[h]
\begin{subfigure}{0.4\textwidth}
\includegraphics[scale=0.35]{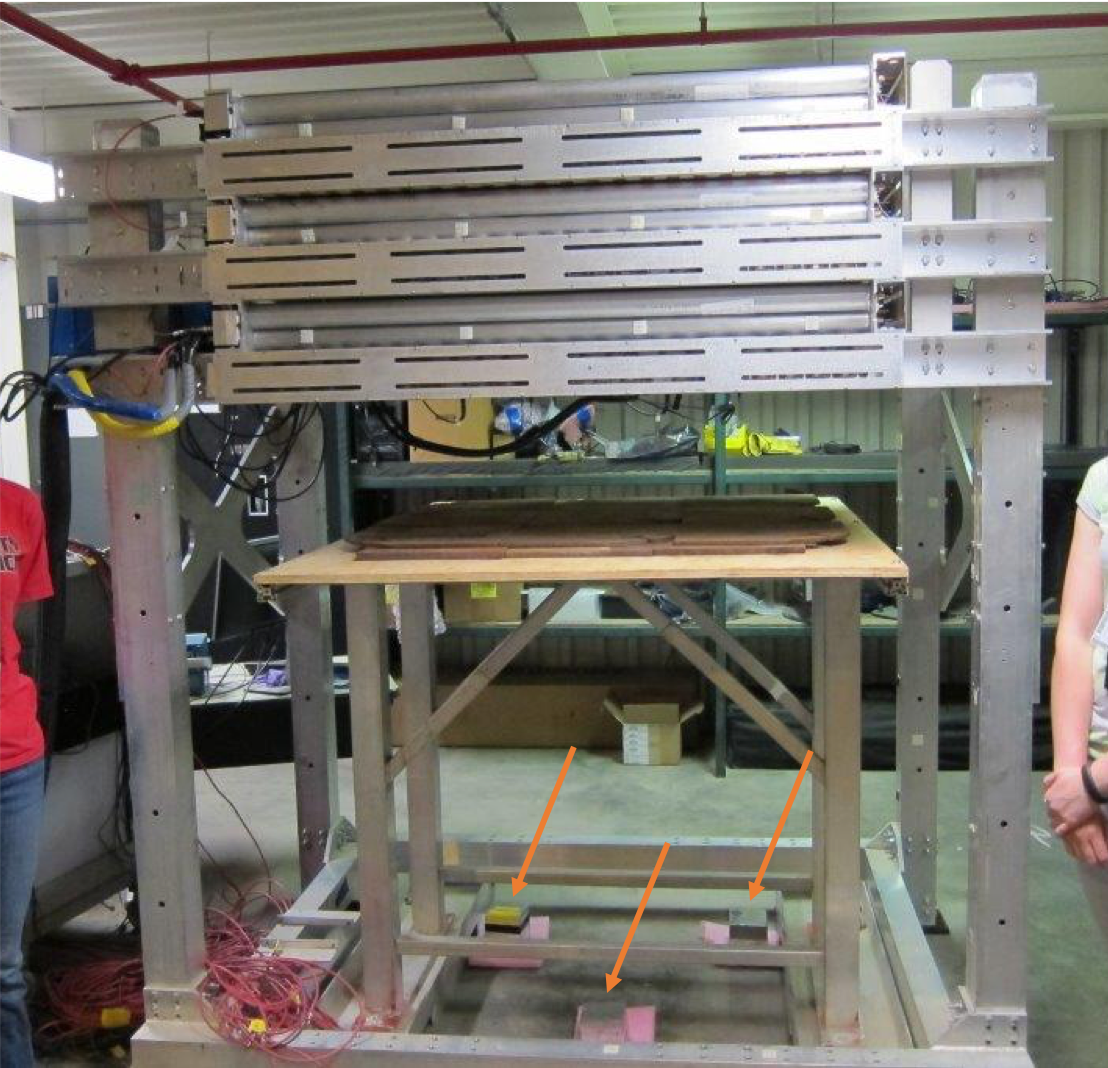}
\label{fig:lightened}
\end{subfigure}
\begin{subfigure}{0.4\textwidth}
\includegraphics[scale=0.5]{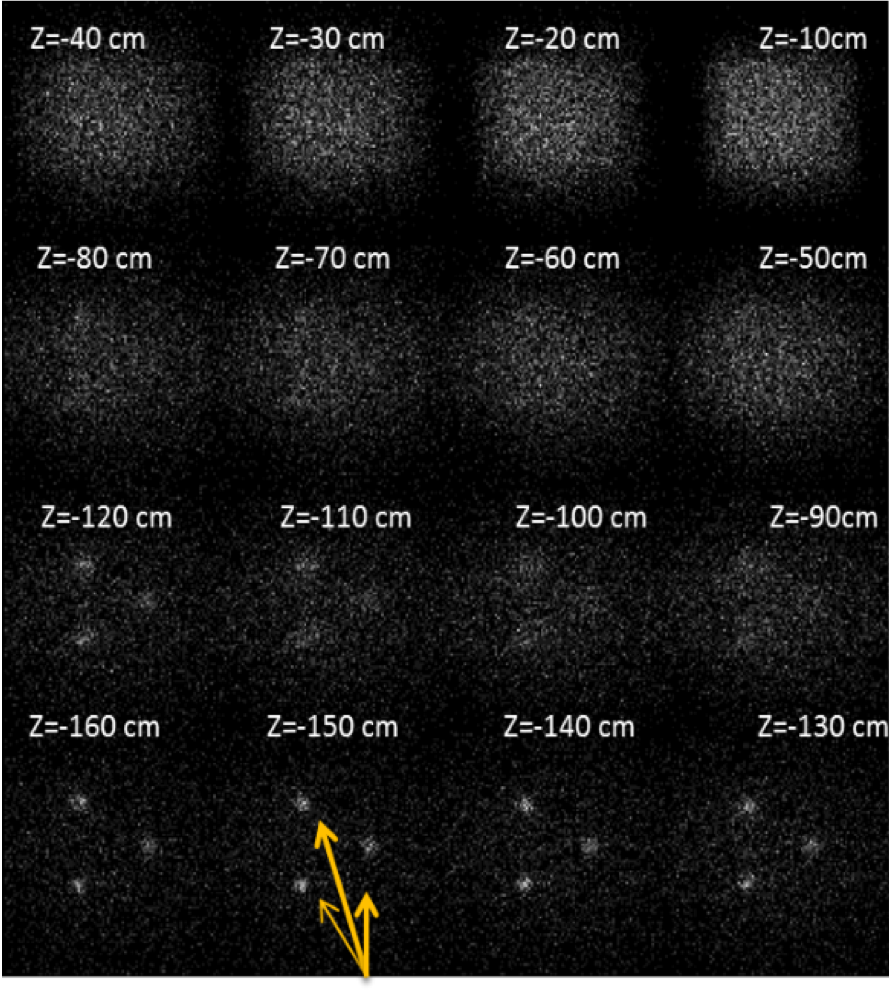}
\label{fig:fig4}
\end{subfigure}
\caption{Left: Setup of the three 20kg U cubes under a muon tracker. Right: The image plane projected to various distances under the muon tracker.  The three distinct neutron emission sources come into focus near $z$=-150cm. }
\label{fig:four}
\end{figure}

This technique was demonstrated at LANL using three 20kg uranium cubes as test objects.  Two of the cubes consisted of depleted uranium, and the other was 19.9$\%$ enriched in $^{235}$U.  Due to the low criticality of these test objects, there was very little neutron multiplication, so the depleted and low-enriched cubes give very similar neutron responses to captured muons.  A single muon tracking detector was placed 1.5m above the uranium cubes, with 2.5cm of steel in between the tracker and cubes to represent an approximation of some shielding.  Stilbene crystal neutron detectors were placed 1.5m away on the floor (see the left panel of Fig.\ref{fig:four} for a picture of the setup).  Data was collected for 24 hours, and the resulting images projected to various distances below the detector are shown in the right panel of Fig.\ref{fig:four}.  The three distinct neutron emission sources come into focus near 150cm below the detector, as expected.  While the time scale for these tests on surrogate material was significant, measurements of actual HEU configurations with significant neutronic gain could dramatically increase the muon-induced neutron emission and thereby decrease the necessary count times.

In a potential warhead verification scenario, such a system could be placed over a closed missile hatch on a submarine, since muons and fast neutrons could penetrate the submarine skin and missile hatch.  This would allow warheads to be counted without requiring any visual access to the warheads or any dismantling or handling of the missile.  Tests on significant quantities of fissile material, as well as a more detailed study of the limits of information obtained with this method, need to be pursued to fully evaluate the technique.

\section{Conclusions}

Cosmic ray muons are passive, penetrating, and freely available everywhere on the Earth's surface.  They provide a radiographic probe that is well suited to solve a variety of difficult imaging problems without the need for any artificial radiation sources.  Further research on applications in nuclear security is underway. 

\section*{Acknowledgments}

This work was supported by the National Nuclear Security Administration's Office of Defense Nuclear Nonproliferation Research and Development, US Defense Threat Reduction Agency, and Los Alamos National Laboratory Lab Directed Research and Development.  This document is released under LA-UR-18-25470.

\bibliography{main}   

\end{document}